\documentclass[12pt]{article}

\usepackage[utf8]{inputenc}
\usepackage[
    twoside,
    top=1in,
    bottom=0.75in,
    inner=0.75in,
    outer=0.75in
]{geometry}
\usepackage{titling}
\usepackage{graphicx}
\usepackage{hyperref}

\title{How do Retail Stores Affect Pedestrian Walking Speed: An Empirical Observation}
\author{Danrui Li}
\date{(Prior to peer review)}

\begin{document}

\markboth{\theauthor}{\thetitle}

\maketitle

\begin{abstract}
    Pedestrian studies in retail areas are critical for comfort and convenience in transportation facility designs. But existing literature lacks detailed empirical observations that focus on pedestrian speed variations and their mechanisms in front of stores. This paper bridges this gap by analyzing 1193 pedestrian trajectories in front of a convenience store located in a metro station. The results show that the store imposes a non-uniform slowing effect on the pedestrian flow. The spatial distribution and the lower walking speed of consumers and gazing pedestrians jointly contribute to such an effect while avoiding behaviors between pedestrians play little role. The findings complement the existing empirical observations and lay a foundation for realistic pedestrian modeling in retail areas.
\end{abstract}

\section{Introduction}
\label{sec:1}

Pedestrian dynamics in public spaces have attained great concern in recent years for their contributions to the planning of transportation facilities, shopping malls, and so on~\cite{hoogendoorn2004}. Amongst all its applications, retail areas in transportation hubs, where architectural designers have to balance between the convenience of users and the safety of pedestrian traffic through store arrangements and facility dimensioning~\cite{hanseler}, play critical roles in the comfort and convenience of city life for their huge usages. Therefore, to better assist related designs or management, it is of great value to develop theories and models that reveal how stores affect pedestrian flows in transportation facilities.

Such a question can be categorized as the study of pedestrian interactions with attractions, where previous literature focused on the following three topics: how do pedestrians make route choices to visit attractions~\cite{borgers1986, bovy2004}, the mechanism of attention shifting when attractions exist~\cite{saunders2004, chen2011, gallup2012, wang2014}, and the locomotion models when interacting with attractions~\cite{helbing95, kwak2013, wangthesis}. To better explore the third topic, it is important to evaluate to what extent existing models fit real-world behaviors in terms of speed and direction variations. But few of the previous studies did so in a comprehensive way.

The problem above may result from the lack of detailedness in related empirical studies. It has been known that aggregated pedestrian speed in retail environments is lowered by stores~\cite{rastogi2013, zacharias2021, alazzawi}. However, little is known about whether such effects are distributed uniformly inside walking facilities, which will be discussed in this paper later. In addition, we don’t know how stores alter pedestrian behaviors that lead to the slowing effect in retail areas. While a traditional assumption may attribute the slowing effect to avoiding behaviors in crossing flows between consumers and normal pedestrians, recent studies in cognitive science experiments~\cite{alyahya,patel} offered an alternative perspective, claiming that pedestrians with cognitive tasks tend to slow their walking speed. Since consumers and gazing pedestrians in retail areas are also pedestrians with cognitive tasks, this paper will compare the impacts of traditional factors (avoiding behaviors) and those of cognitive factors (consuming and gazing behaviors) on the slowing effects. 

Therefore, this paper aims to enrich the empirical studies in pedestrian dynamics in retail areas by investigating to what extent and how stores affect the walking speed of bi-directional pedestrian flows. Using trajectories collected from field observations, two questions are explored as follows: (1) the walking speed variations inside the walking facility, and (2) the main behavior that causes the speed variation.

\section{Method}
\label{sec:2}

\paragraph{Data Collection and Labelling}

An underground bi-directional corridor in a metro station was selected as the field observation site. This corridor only serves metro passengers transferring between different lines. Hence, pedestrian movement on strategic and tactical layers can be ignored. One single convenience store is located on one side of this corridor with no other stores, obstacles, or attractors around. 

\begin{figure}[b]
\centering
\includegraphics[width=6cm]{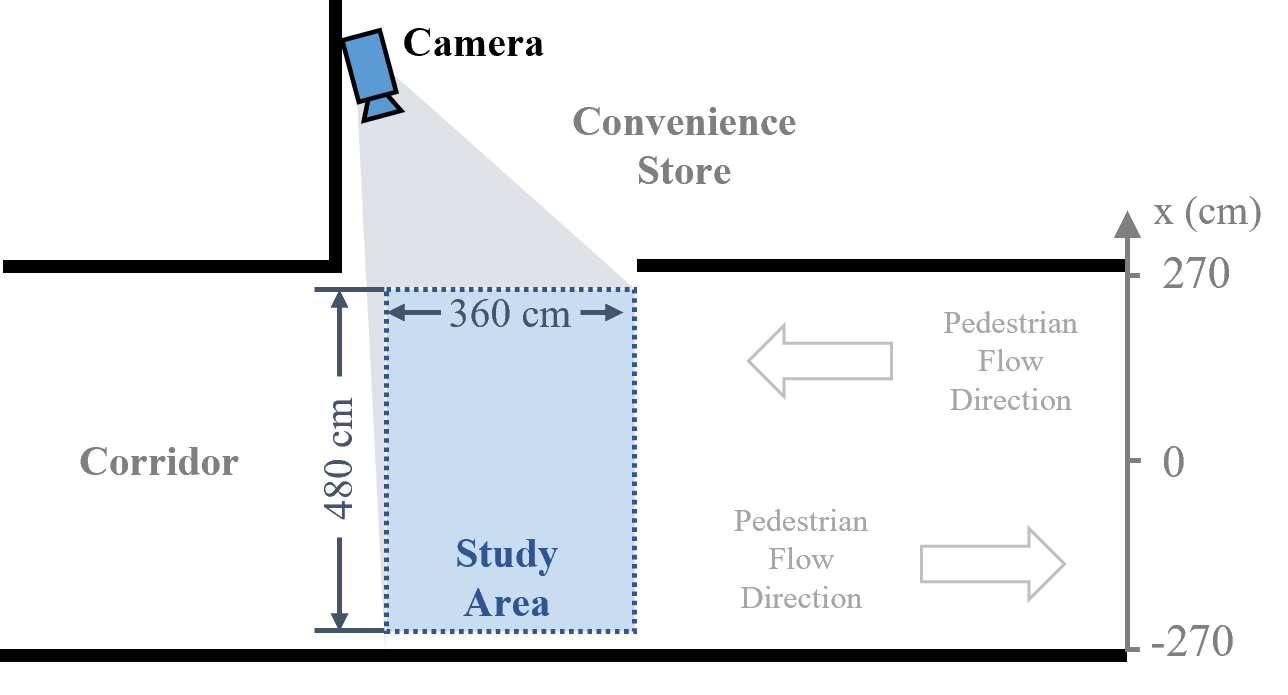}
\caption{Field observation site and the camera installation. The study area covered the front of a convenience store. And the origin of the x-axis was set to the middle point of the corridor section line. }
\label{fig:install}
\end{figure}

One camera was installed inside the store entrance without obstructing pedestrians (see Fig.\ref{fig:install}). Through 45-minute-long video recordings in off-peak hours during the daytime, the movements of 1193 pedestrians are captured in front of the store.

Since pedestrian occlusions were so commonly seen in the video recordings due to the non-orthogonal camera installation, based on the approaches from previous studies~\cite{gallup2012}, pedestrian positions are labeled manually for every 15 frames (0.5 seconds) by estimating the points on the ground directly below the bodies. 

Besides, pedestrian behaviors are also tagged as below:

\begin{itemize}

\item Consuming: Walking in or out of the store is tagged as a consuming behavior.

\item Gazing: Gazing at the store for more than 1.5 seconds. The gazing behaviors are labeled by manual judgments for every 15 frames (0.5 seconds). Consumers are not tagged as gazing pedestrians.

\item Commuting: Pedestrians not tagged as consuming or gazing pedestrians are tagged as commuting.
\item Avoiding: Pedestrians (including consumers and gazing pedestrians) with headways less than 180 cm and alignments less than 1.5 are tagged as avoiding pedestrians. The definition of headways and alignments are in accordance with previous studies by Duives~\cite{duives2016}. This tag can coexist with the first three tags.
\end{itemize}

\paragraph{Average walking speed}

The study area is split into $16$ lanes in equal width by x coordinates. The areas that are less than $30$ cm from the walls are excluded due to the scarcity of data. The walking speed for each lane is an average of all individual-aggregated data points falling into this lane:

\begin{equation}
\mathcal{V}_k = \frac{1}{|P_k|}\sum_{p \in P_k} (\frac{1}{N_{k,p}} \sum_{i=1}^{N_{k,p}} |v_i|)
\end{equation}

where $\mathcal{V}_k$ is the average walking speed of lane $k$. $P_k$ contains all pedestrians whose data points fall into lane $k$. $N_{k,p}$ represents the total number of data points of pedestrian $p$ on lane $k$. And $v_i$ represents the velocity of data point $i$.

In this way, speed variation can be observed by comparing the average speed of pedestrians on different lanes. Since previous studies~\cite{zanlungo2014}  revealed that speed can be a function of lateral position (x coordinate) in bi-directional corridors without stores, speed differences between the two sides of the corridor are also compared to eliminate such effects: 

\begin{equation}
\Delta \mathcal{V}_k = \mathcal{V}_{k} - \mathcal{V}_{k'}
\end{equation}

where $\mathcal{V}_k$ and $\mathcal{V}_{k'}$ are the average walking speed of lane $k$  and lane $k'$. Lane $k$ covers all data points whose x coordinates fall into $[2kl-2l, 2kl)$. And the coverage of lane $k'$ is symmetrical about the center of the corridor, i.e. $[-2kl+2l, -2kl)$. In this paper $l = 15cm$. So a set of $\Delta \mathcal{V}_k$ for $8$ lane pairs can be derived.

\paragraph{Compare impacts of different behaviors}

Three factors (avoiding, gazing, and consuming behaviors) leading to the asymmetrical speed function will be compared by observing the changes in $\Delta \mathcal{V}_k$ when the corresponding tagged pedestrians are removed from the sample. The assumption is: if the most influential groups are removed, the $\Delta \mathcal{V}_k$ in all lanes will decrease to the largest extent. Ideally, the $\Delta \mathcal{V}_k$ in all lanes will be $0$ if all factors for the slowing effects are found. 

\section{Results}
\label{sec:3}
\paragraph{Non-uniform Slowing Effect by the Store}

\begin{figure}[hbt]
\centering
\includegraphics[width=11.5cm]{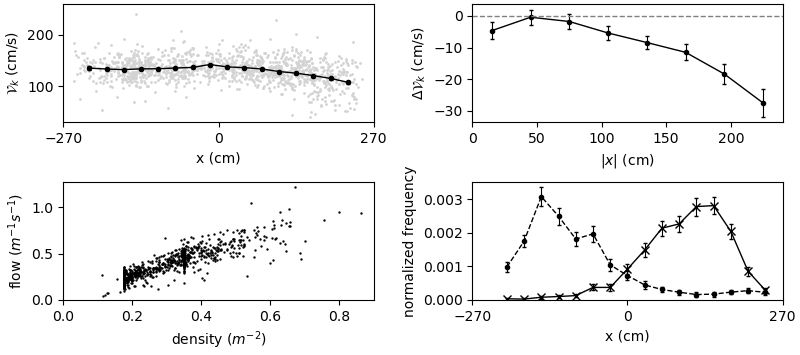}
\caption{Top left: The average walking speed $\mathcal{V}_k$ as a function of the position in the cross-section of the corridor (in black lines). Scatter plot of the average x coordinate and the average speed for each individual (in grey dots). Top right: The walking speed difference $\Delta \mathcal{V}_k$ between the two sides of the corridor as a function of the distance to the central line ($x=0$). Error bars = standard errors of means. Bottom left: Scatter plot of the average adjacent density and the average adjacent flow rate for each individual (in grey dots). Bottom right: The histogram of the pedestrian distribution along the x-axis. The two lines represent pedestrians in reversed directions.}
\label{fig:overview}
\end{figure}

The result (see Fig.~\ref{fig:overview}) shows that the walking speed distribution is not symmetrical about the central line of the corridor. As shown in the top left figure, the average speed on the store side ($x > 0$) is lower than that on the opposite side ($x < 0$). And such an asymmetricity increases with the distance to the central line.

This effect can be attributed to the store for the following reasons: as shown in the last two figures in Fig.~\ref{fig:overview}, the pedestrian flow is diluted and the pedestrian densities of the two directions are similar. So we can assume that the speed function should be symmetrical about the central line if there are no stores around. Now that asymmetricity emerges, we can conclude that such asymmetricity results from the store.

\paragraph{Consuming and Gazing Behaviors are Dominant Factors}

\begin{figure}[htb]

\centering
\includegraphics[width=11cm]{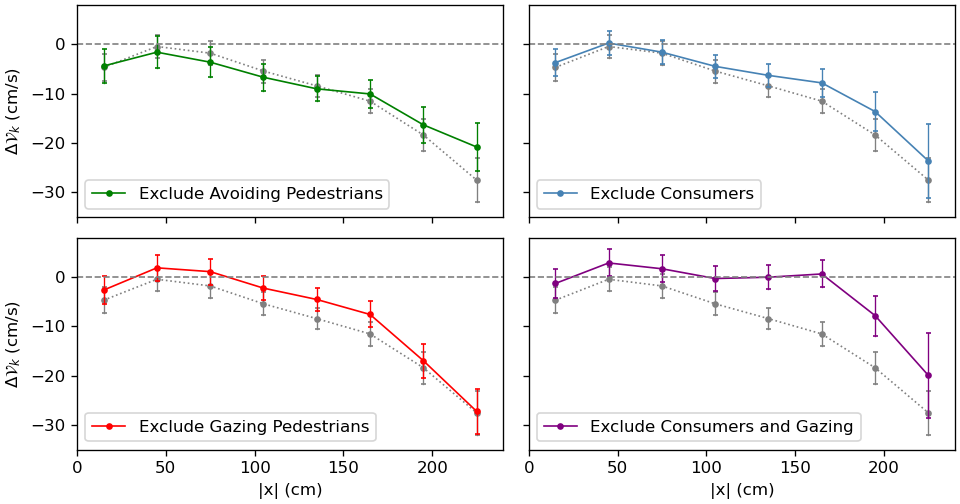}
\caption{The walking speed difference between the two sides of the corridor as a function of the distance to the central line. Grey dotted lines represent the differences for all pedestrians. Colored lines represent those for certain pedestrian groups detailed in the corresponding legends. Error bars = standard errors of means.}
\label{fig:spd}
\end{figure}

The results show that avoiding behaviors actually impose little impact on diluted flows. But consuming and gazing behaviors can jointly explain most of the slowing effects. As shown in Fig.~\ref{fig:spd}, removing pedestrians with avoiding behaviors will only change the asymmetricity of the speed distribution when $|x| > 150$, indicating its effects are limited to the entrance of the store. Gazing pedestrians, however, impose their effects on $|x| \in (50, 150)$, which is the center of the pedestrian flow. When both consumers and gazing pedestrians are removed, the remaining pedestrians show little difference in speed between the two sides, indicating that those two behaviors are the main reasons for the asymmetric slowing effects.

\begin{figure}[hbt]
\centering
\includegraphics[width=11cm]{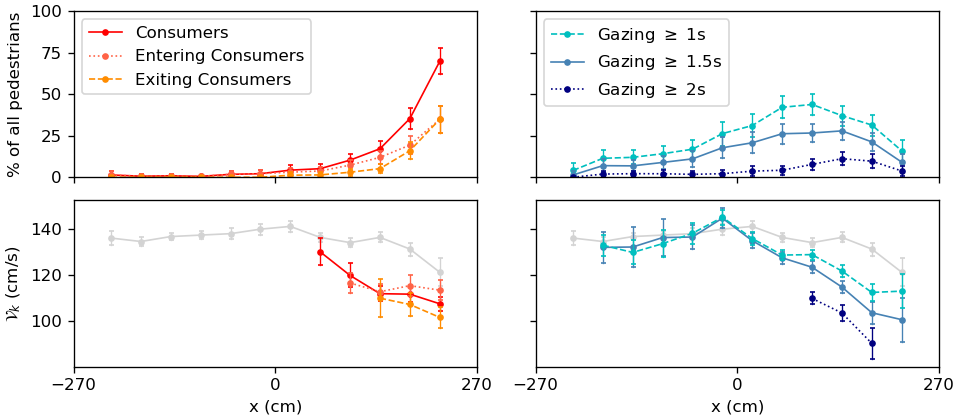}
\caption{Top row: the proportion to all pedestrians as a function of x coordinate (lateral distance to the store entrance) for consumers (top left) and gazing pedestrians (top right). Bottom row: the average walking speed as a function of x coordinate for consumers (bottom left) and gazing pedestrians (bottom right). Grey solid lines represent the walking speed of commuters. Error bars = standard errors of means.}
\label{fig:hetero}
\end{figure}

Finally, how consuming and gazing behaviors contribute to the slowing effect is explored in detail (see Fig. \ref{fig:hetero}). The results show that pedestrians with such behaviors are distributed unevenly across the section of the corridor. While consumers are more likely to show up at the entrance of the store, gazing pedestrians are keeping a certain distance away from the entrance.

Besides, their walking speeds are also the functions of pedestrian lateral positions and internal attributes. Both groups walk more slowly when they get closer to the store, while normal pedestrians are hardly affected by this. Exiting consumers walk more slowly than those who enter the store. A larger speed decrease is observed in gazing pedestrians if they are gazing at the store for a longer time, which is in line with controlled experiment results in previous studies~\cite{alyahya, patel}.

\section{Discussion and Conclusion}
\label{sec:4}

In summary, this paper explores the pedestrian walking speed variation in front of a convenience store and how three factors (avoiding, consuming, and gazing behaviors) influence it by analyzing 1192 pedestrian trajectories. The findings are listed below:

Complementing previous research~\cite{rastogi2013, zacharias2021, alazzawi}, this paper reveals that the slowing effect is distributed unevenly within the walking facilities. Its impacts are much stronger when pedestrians are close to the store entrance compared to the cases when they are in the middle of the walking facility. 
In the future, such findings should be testified in more diversified walking environments. For instance, slowing effects may differ when pedestrians are interacting with different store types (such as convenience stores, restaurants, and beverage takeaway stores). Besides, it remains unknown whether the intensity of the slowing effects is solely influenced by the absolute distances to the stores or also affected by the widths of the walking facilities. In addition, the speed variations in the longitudinal sections of the walking facilities should be studied with richer data.

Among all three factors (avoiding, consuming, and gazing behaviors), the slowing effect is mainly contributed by the latter two. Avoiding behaviors, which are often seen in high-density crossing flows, impose little impact in this paper. Such findings indicate that traditional pedestrian dynamics theories that focus on avoiding behaviors cannot fully explain the phenomena in retail areas where pedestrian densities can be much lower. Therefore, in future works, we need to pay attention to the modeling of certain pedestrian behaviors such as consuming and gazing to achieve more accurate simulations in retail areas. Furthermore, as pedestrian flows get denser, such impact distributions may change, which needs further studies.

This paper reveals the correlation between walking speed, lateral position, and gazing behaviors, proving that the interference between gazing and walking can be not only observed in controlled experiments~\cite{alyahya, patel} but also in field studies. It also suggests a visually-guided locomotion theory may be suitable for gazing pedestrians. Such a theory can be further explored in the future.

The findings above also provide additional perspectives to evaluate existing locomotion models~\cite{saunders2004,wang2014,kwak2013}. For instance, quantitative evaluations of pedestrian speed variation can be carried out to compare or improve those models, while they are only qualitatively assessed in previous literature. 

\section{Acknowledgement}

The author would like to thank Professor Yu Zhuang from Tongji University for his supervision, and Dr. Lingzhu Zhang for her research suggestions. And thank Liuqing Wu, Junkai Wang, Ruixiang Lin, Wenwen He, and Xiyan Yang for their assistance in data collection.


\end{document}